\documentclass[12pt]{article}
\usepackage[xdvi]{graphicx}
\usepackage{amssymb}
\newcommand{\beq}{\begin{equation}}
\newcommand{\eeq}{\end{equation}}
\newcommand{\bea}{\begin{eqnarray}}
\newcommand{\eea}{\end{eqnarray}}

\def\e2sig{e^{-2r\sigma}}

\newcommand{\ccr}{c_{\textrm{\scriptsize cr}}}

\begin{document}

\begin{titlepage}
\begin{flushright}
KOBE-TH-06-07 \\
TIT/HEP-560 \\
HIP-2006-49/TH \\
hep-th/0612071
\end{flushright}
\vspace*{2cm}
\begin{center}{\large\bf  
Radius stabilization by constant boundary superpotentials
in warped space}

\end{center}
\vspace{1cm}
\begin{center}
{\bf Nobuhito Maru}$^{(a)}$
\footnote{E-mail: maru@people.kobe-u.ac.jp},
{\bf Norisuke Sakai}$^{(b)}$
\footnote{E-mail: nsakai@th.phys.titech.ac.jp},
~and~~{\bf Nobuhiro Uekusa}$^{(c)}$
\footnote{E-mail: nobuhiro.uekusa@helsinki.fi}
\end{center}
\vspace{0.2cm}
\begin{center}
${}^{(a)}$ {\it Department of Physics, Kobe University, 
Kobe 657-8501, Japan}
\\[0.2cm]
${}^{(b)}$ {\it Department of Physics, Tokyo Institute of Technology, 
Tokyo 152-8551, Japan}
\\[0.2cm]
${}^{(c)}$ {\it High Energy Physics Division, 
Department of Physical Sciences, University of Helsinki,  
and Helsinki Institute of Physics, 
P.O. Box 64, FIN-00014 Helsinki, Finland}
\end{center}
\vspace{1cm}
\begin{abstract}

A warped space model with a constant boundary 
superpotential has been an efficient model 
both to break supersymmetry and to stabilize the radius, 
when 
hypermultiplet, compensator and radion multiplet are 
taken into account. 
In such a model of the radius stabilization, 
the radion and moduli masses, the gravitino mass and 
the induced soft masses are studied.
We find that a lighter physical mode composed of 
the radion and the moduli can have mass of the order of 
a TeV and that the gravitino mass can be of the order of 
10$^7$ GeV. 
It is also shown that soft mass induced by the 
anomaly mediation can be of the order of 100GeV and can 
be dominant compared to that mediated by bulk fields. 
Localized F terms and D terms are discussed as candidates of 
cancelling the cosmological constant. 
We find that there is no flavor changing neutral current problem 
in a wide range of parameters.

\end{abstract}
\end{titlepage}
\newpage
\renewcommand{\theequation}{\thesection.\arabic{equation}} 
\setcounter{equation}{0}
\section{Introduction}
Supersymmetry is a well-motivated 
extension to the Standard Model, which plays a crucial 
role in solving the gauge hierarchy problem \cite{Dimopoulos:1981zb}. 
Extra dimensions with flat space \cite{Arkani-Hamed:1998rs}
or with the 
warped space \cite{Randall:1999ee} are also an alternative solution 
to the gauge hierarchy problem. 
Considering both ingredients is natural in the context 
of the string theory and is often taken as the starting 
point in the phenomenological model of the brane world 
scenarios.

There is another motivation to consider the brane world 
scenario in the context of supersymmetry breaking mediation 
in supergravity (SUGRA) \cite{Randall:1998uk}.
In 4D SUGRA, once supersymmetry is broken in the hidden 
sector, its breaking effects can be mediated to the visible 
sector through the Planck suppressed interactions and 
soft masses can be of the order of the gravitino mass. 
Although the soft supersymmetry breaking masses are 
severely constrained to be almost flavor diagonal by 
experiments, there is no symmetry reason for such a flavor 
structure in 4D SUGRA. 
Therefore 4D SUGRA models generically suffer from the 
flavor-changing-neutral-current (FCNC) problem. 
If the two sectors are separated each other along the 
direction of extra 
dimensions \cite{Randall:1998uk,Luty:1999cz}, 
interactions  between the visible and the hidden 
sectors are naturally suppressed.  
In this setup, soft supersymmetry breaking terms in the 
visible sector are generated through the superconformal 
anomaly (anomaly mediation) and the resultant mass 
spectrum is found to be flavor-blind, namely 
there is no FCNC problem. 
If anomaly mediation dominates,
tree level gravity contributions should be suppressed
(for review see \cite{Luty:2005sn}).
4D realizations of this idea have also been discussed 
\cite{Luty:2001jh, Luty:2001zv, Nomura:2005rj}.
More recently, attempts to probe anomaly mediation
through TeV scale measurements have been discussed
\cite{Cohen:2006qc, Choi:2007ka}.

In the brane world scenario, there is an important issue 
of radius stabilization. 
In order for the scenario to be phenomenologically viable, 
the compactification radius should be stabilized. 
In the previous paper~\cite{Maru:2006id},
we gave a model of radius stabilization.
Here we investigated supersymmetry-breaking 
effects caused by constant 
(field independent) superpotentials localized 
at fixed points in the supersymmetric Randall-Sundrum 
model. 
By taking into account the hypermultiplet, the 
compensating multiplet and the radion multiplet, 
we have shown that
the radius is stabilized by the presence of 
the constant boundary superpotentials\footnote{%
Supersymmetry breaking by a (bulk) constant superpotential 
related to Scherk-Schwarz supersymmetry breaking has been discussed 
in the literature \cite{Bagger:2003vc,Bagger:2003fy}
in which it has been shown that
the radius is not stabilized in the Randall-Sundrum model only with
gravity multiplet.
Such a destabilization was shown also in \cite{Maru:2005qx}.
For other discussions on stabilization,
see \cite{Maru:2003mq,Eto:2004yk,Abe:2004ar,Abe:2005ac,Correia:2006pj,Abe:2006eg}, for example.}.
%
However there remain to be examined 
a number of important phenomenological issues, 
such as FCNC and 
masses of superparticles which necessarily appear in the 
model.

In this paper, we calculate the radion and moduli masses,
the gravitino mass and the induced soft masses in 
the model given by Ref.\cite{Maru:2006id}.
We find that a lighter physical mode composed of 
the radion and the moduli can have masses of the order of 
a TeV and that the gravitino mass can be of the order of 
10$^7$ GeV. 
It is also shown that induced mass mediated by 
anomaly can be of the order of 100GeV and can be dominant 
compared to that mediated by bulk fields. 
We discuss cancelling the cosmological constant with
F terms and D terms localized at $y=0$ so as not to 
affect radius stabilization.
We find that there is no FCNC problem in a wide range of 
parameters.

The paper is organized as follows. 
The model is introduced in Sec.\ref{sc:model}. 
The radion and moduli masses are calculated 
in Sec.\ref{sc:revisedkinadd}.
Soft masses induced by anomaly mediation are obtained 
in Sec.\ref{sc:08082006-3anomaly}.
In Sec.\ref{sc:w0model-4}
we give Kaluza-Klein masses of hyperscalar and gravitino
and then estimate soft masses 
induced by mediation of all Kaluza-Klein modes.
Section~\ref{sc:08082006-3} is devoted to 
discussion about localized F terms.
Soft masses induced by localized F terms 
are calculated. 
In Sec.~\ref{sc:FIs}, cancellation of the cosmological constant
by D terms is analyzed.
Conclusion is given in Sec.\ref{sc:conclusion}.
The details of calculations of mass spectrum 
of hyperscalar and gravitino are shown in Appendix.

\setcounter{equation}{0}
\section{Model}\label{sc:model}
We consider a five-dimensional supersymmetric model of 
a single hypermultiplet on the 
Randall-Sundrum 
background, 
whose metric is 
\bea
ds^2 = e^{-2R\sigma}\eta_{\mu\nu}dx^\mu dx^\nu +R^2 dy^2, 
\quad 
\sigma(y)\equiv k|y|, 
\eea
where $\eta_{\mu\nu}={\rm diag.}(-1,+1,+1,+1)$, 
$R$ is the radius of $S^1$ of the orbifold $S^1/Z_2$, 
$k$ is the $AdS_5$ curvature scale, and the angle of $S^1$ 
is denoted by $y(0 \le y \le \pi)$. 
In terms of superfields for four manifest supersymmetry, 
our Lagrangian reads \cite{Maru:2006id,Marti:2001iw}
\bea
{\cal L}_5 &=& \int d^4 \theta 
\frac{1}{2} \varphi^\dag \varphi (T+T^\dag) 
e^{-(T+T^\dag)\sigma}
(\Phi^\dag \Phi + \Phi^c \Phi^{c\dag} - 6M_5^3) 
\nonumber \\
&& + \int d^2 \theta 
\left[
\varphi^3 e^{-3T \sigma} \left\{
\Phi^c \left[
\partial_y - \left( \frac{3}{2} - c \right)T \sigma' 
\right] \Phi + W_b
\right\} + {\rm h.c.}
\right] ,
\label{lagrangian}
\eea
where the compensator chiral supermultiplet $\varphi$ 
(of supergravity), and the radion chiral supermultiplet 
$T$ are denoted as~\footnote{In the present paper,
we are interested in the (real part of) radion and the 
hyperscalars.
If we also consider the imaginary part 
of the radion supermultiplet naively in the present model, 
we find it to be massless since the potential depends only 
on the real part of the radion. 
Therefore, we may require some other mechanisms which give 
the imaginary part a mass or suppress its coupling to the 
standard model fields in order to make our model fully 
compatible with experiments. 
} 
\bea
\varphi = 1 + \theta^2 F_{\varphi}, 
\qquad 
T=R + \theta^2 F_T , 
\eea
respectively, and the chiral supermultiplets representing 
the hypermultiplet is denoted as 
$\Phi, \Phi^c$.
The $Z_2$ parity is assigned to be even (odd) for 
$\Phi (\Phi^c)$. 
The derivative with respect to $y$ is denoted by $'$, 
such as $\sigma'\equiv d\sigma/dy$. 
The five-dimensional Planck mass is denoted as $M_5$. 
Here we consider a model with a constant (field independent) 
superpotential localized at the fixed point $y=0$ 
\bea
W_b \equiv 2M_5^3 w_0 \delta(y) ,
\label{eq:boundary_pot}
\eea
where $w_{0}$ is a dimensionless constant.
As far as the analysis in Sec.~\ref{sc:revisedkinadd} is 
concerned, our Lagrangian turns out to be equivalent, 
with the replacement $w_0\to (3/2)w_0$ in 
(\ref{eq:boundary_pot}), to the Lagrangian
\footnote{%
The Lagrangian is explicitly written as (2.10) in 
Ref.\cite{Maru:2006id}. 
For other aspects of the Lagrangian, it is quite 
complicated to derive the Lagrangian from the 5D SUGRA 
and to identify the radion. 
Although there have been some attempts to understand  
the radion 
\cite{PaccettiCorreia:2004ri,Abe:2004ar,Abe:2005ac,Abe:2006eg} 
and its imaginary part \cite{Correia:2006pj} in 5D SUGRA, 
we leave it as a future work. 
Analyses on anomaly mediation and gravitino are given 
by following the well-known formulations in 
Refs.\cite{Luty:2002ff} and \cite{Gherghetta:2000qt}, 
respectively. 
We study localized F and D terms based on 4D SUGRA as they are
confined on the brane at $y=0$ where the warp factor is trivial.}
proposed in Ref.~\cite{Abe:2004ar} based on a more 
accurate treatment of the radion superfield using 5D SUGRA.

As shown in Ref.\cite{Maru:2006id},
the background solutions for the scalar components 
at the leading order of $w_0$ are given by
\begin{eqnarray}
\phi(y)=N_2\exp\left[\left({3\over 2}-c\right)R\sigma\right], 
 \label{phin2} 
\end{eqnarray}
\begin{eqnarray}
\phi^c(y)=\hat{\epsilon}(y)
   \left({\phi^\dagger\phi\over 6M_5^3}-1\right)^{-1}
 \left({\phi^\dagger\phi\over 6M_5^3}\right)^{{5/2-c\over 3-2c}}
   \left[c_1 +c_2 
 \left({\phi^\dagger\phi\over 6M_5^3}\right)^{-{1-2c\over 3-2c}}
\left({\phi^\dagger\phi\over 6M_5^3}+{2\over 1-2c}\right)\right]
\label{phichic}
\end{eqnarray}
where $c\neq 1/2, 3/2$, and 
\begin{eqnarray}
\hat{\epsilon}(y) \equiv 
\left\{
\begin{array}{cc}
+1, & 0<y<\pi \\
-1, & -\pi<y<0
\end{array}
\right.
. 
\label{eq:sign_function}
\end{eqnarray}
The solution contains three complex integration constants: 
$c_1, c_2$ are the coefficients of two independent 
solutions for $\phi^c$, and the overall complex constant 
$N_2$ for the flat direction $\phi$. 
Two out of these three complex integration constants 
are determined by the boundary conditions. 
The single remaining constant (which we choose as $N_2$) 
is determined through the minimization of the potential 
(stabilization). 
With these backgrounds, the potential is obtained as 
\bea
 V
&=& {3M_5^3 k w_0^2\over 2}\bigg\{
  \frac{-2(1-2c)}{(1-2c)(e^{2Rk\pi}-1)\hat{N} 
+2(e^{(2c-1)Rk\pi}-1)}
\hat{N}^{4-2c-\frac{1}{3-2c}} \nonumber \\
&&+\frac{\hat{N}}{1-\hat{N}}
\left( -4c^2+12c-6 +\frac{3-2c}{3(1-\hat{N})}
\right)
  \bigg\}. 
\label{potentialwp0}
\eea
where $\hat{N}\equiv |N_2|^2/(6M_5^3)$.
There is a unique nontrivial minimum with a finite value 
of both the radius $R$ and the overall constant 
$N_2$ for the direction $\phi$, provided 
$c < c_{\rm cr}$ with 
\begin{eqnarray}
  c_{\textrm{\scriptsize cr}}\equiv
    {17-\sqrt{109}\over 12} .
\end{eqnarray}
The stationary conditions are solved with
\begin{eqnarray}
  \hat{N}=e^{-(3-2c)Rk\pi} . \label{stanrk}
\end{eqnarray}
For $c=\ccr-\Delta c$ with a small $\Delta c$,
the stationary point at 
the leading order of $\Delta c$ is obtained as 
\begin{eqnarray}
  R&\approx&
    {-1\over \left[2(1-\ccr)(3-2\ccr)+1\right]k\pi}
      \ln \left[
     {(3-2\ccr)\left({17\over 3}-4\ccr\right)
     \over 2(2\ccr-1)\left(2-\ccr-{1-\ccr\over 3-2\ccr}\right)}
   \Delta c\right] 
\nonumber \\ 
 &\approx& {1\over 10k}\left(\ln {1\over \Delta c}-3.4\right) ,
\label{Rdeltac} 
\end{eqnarray}
which means that the radius is stabilized with the size 
of $R>1/k$ 
for $\Delta c<10^{-6}$. 
In our setup, the gauge hierarchy problem can be solved 
by supersymmetry, which is unbroken in the limit of 
vanishing boundary superpotential $W_b$. 
Therefore we do not necessarily need an enormous 
hierarchy by the warp factor, such as 
$e^{-\pi Rk}\sim 10^{-16}$. 

At the stationary point the potential becomes 
\begin{eqnarray}
  V&\approx& -10^{37}(kw_0)^2(\Delta c)^{1.2} .
  \label{eq:pst}
\end{eqnarray}
This negative 
vacuum energy 
may be shifted by contributions of 
other sources for supersymmetry breaking.
As we will show in Sec~\ref{sc:08082006-3},
localized F term is a candidate of such a source 
and can make the energy at the stationary point  
approximately zero (or a tiny positive cosmological 
constant within observational bound).
With this cancellation of the cosmological constant,
we will work with 4D flat background 
rather than $AdS_4$ background\footnote{In $AdS_4$ background, 
supersymmetry breaking by constant boundary superpotential 
is closely related to Scherk-Schwarz supersymmetry breaking. 
In the case of Scherk-Schwarz supersymmetry breaking 
in $AdS_4$ the radion potential and soft masses were given 
in Ref.\cite{Katz:2006mv}.}.

\setcounter{equation}{0}
\section{Radion and moduli masses}\label{sc:revisedkinadd}

In this section, we calculate the masses for
the quantum fluctuations of the radion and moduli 
in our model by computing the piece of Lagrangian 
quadratic in fluctuations consisting of kinetic terms 
and mass terms 
\begin{eqnarray}
 {\cal L}_{\textrm{\scriptsize quadratic}}=
 {\cal L}_{\textrm{\scriptsize kin}}+
 {\cal L}_{\textrm{\scriptsize mass}}
\end{eqnarray}
We start with the kinetic part of the quantum 
fluctuations in order to find the diagonalized and 
canonically normalized fluctuation fields. 
From Eq.(\ref{lagrangian}), the kinetic Lagrangian is 
written as
\begin{eqnarray}
 {\cal L}_{\textrm{\scriptsize kin}}
&=& \sigma (\partial_\mu R)(\partial^\mu R)e^{-2R\sigma}
  (\phi^\dagger\phi+\phi^c{}^\dagger\phi^c-6M_5^3 )
\nonumber\\
  &&-{1\over 2}(1-2R\sigma)
e^{-2R\sigma}(\partial_\mu R)\partial^\mu
   (|\phi|^2+|\phi^c|^2)
\nonumber\\
 &&- Re^{-2R\sigma}\left[
  (\partial_\mu
  \phi^\dagger)(\partial^\mu \phi)
  + (\partial_\mu
  \phi^c{}^\dagger)(\partial^\mu \phi^c)\right] ,
   \label{lag}
\end{eqnarray}
where we performed a partial integral and dropped 
four-dimensional total derivatives. 
Since the field $\phi^c$ is of higher order of $w_0$, 
we will omit $\phi^c$ in Eq.(\ref{lag}). 
Without loss of generality, we can choose the phase of the 
background classical solution in Eq.(\ref{phin2}) as 
\begin{eqnarray}
N_2=N_2^\dagger. 
\end{eqnarray}
We now introduce quantum fluctuation fields around the 
background classical solution to define the radion 
$\tilde R$ and the moduli field $\tilde{N}_2$ : 
\begin{eqnarray}
R+\tilde{R}, \quad N_2+\tilde{N}_2, \quad 
\tilde{N}_2=\tilde{N}_{2R}+i\tilde{N}_{2I}. 
\end{eqnarray}
Substituting Eq.(\ref{phin2})
into ${\cal L}_{\textrm{\scriptsize kin}}$, 
we find that the 4D kinetic Lagrangian for the quantum 
fluctuations has a mixing between radion $\tilde R$ and 
the real part of the moduli filed $\tilde{N}_{2R}$ 
\begin{eqnarray}
 \int_0^\pi dy ~ {\cal L}_{\textrm{\scriptsize kin}}
 = -(\partial_\mu \tilde{R}, \partial_\mu \tilde{N}_{2R}
)
   \left(\begin{array}{cc}
       f_{11}&f_{12}\\ 
       f_{21}&f_{22}\\
	 \end{array}\right)
    \left(\begin{array}{c}
     \partial^\mu \tilde{R}\\
     \partial^\mu \tilde{N}_{2R}\\
   \end{array}\right) 
-f_{22} \partial_\mu \tilde{N}_{2I} \partial^\mu \tilde{N}_{2I},
\end{eqnarray}
where
\begin{eqnarray}
  f_{11} &\equiv&
  {|N_2|^2\over (1-2c)^3 R^2k}
  e^{(1-2c)Rk\pi}\bigg\{
   -\left({3\over 2}-c\right)\left({1\over 2}+c\right)(1-2c)^2
   (Rk\pi)^2
\nonumber
\\
 &&
  +2(1-2c)(Rk\pi)
  -2+2 e^{-(1-2c)Rk\pi}
  \bigg\}
 +{3M_5^3 \over 2R^2k}
  \bigg\{
  1-e^{-2Rk\pi}(1+2Rk\pi)
   \bigg\} ,
\\
 f_{12} &\equiv& {\pi \over 2}N_2^\dagger
    e^{(1-2c)Rk\pi}
  =  {\pi \over 2}N_2 
    e^{(1-2c)Rk\pi}
  = f_{21} ,
\\
 f_{22}&\equiv&
    {e^{(1-2c)Rk\pi}-1\over (1-2c)k} .
\end{eqnarray}
To transform this Lagrangian into canonical forms,
we can take the following basis 
\begin{eqnarray}
  \overline{N}_{2I}
= \sqrt{2f_{22}} \tilde{N}_{2I}, 
\end{eqnarray}
and 
\begin{eqnarray}
 \left(\begin{array}{c}
  \overline{R}\\
  \overline{N}_{2R}
       \end{array}\right)
 =\left(\begin{array}{cc}
    \sqrt{\lambda_+}& \\
    & \sqrt{\lambda_-} \\
        \end{array}\right)
  \left( \begin{array}{cc}
   \cos\theta &\sin\theta \\
  -\sin\theta &\cos\theta \\  
         \end{array}\right)
  \left(\begin{array}{c}
   \tilde{R}\\
   \tilde{N}_{2R}\\
        \end{array}\right) ,
\end{eqnarray}
where $\lambda_\pm$ and the rotation angles are given by
\begin{eqnarray}
 \lambda_\pm&=&
   f_{11}+f_{22}\pm
   \sqrt{(f_{11}-f_{22})^2+4f_{12}^2} ,
\\
 \tan\theta&=&{1\over 2f_{12}}
   \left\{f_{22}-f_{11}
 +\sqrt{(f_{11}-f_{22})^2+4f_{12}^2}\right\} .
\end{eqnarray}

Let us now evaluate the radion and moduli masses.
The mass terms are given by
\begin{eqnarray}
 \int_0^\pi dy~ {\cal L}_{\textrm{\scriptsize mass}}
 =-{1\over 2}
(\overline{R}, \overline{N}_{2R}) {\cal M}^2
   \left(\begin{array}{c}
    \overline{R} \\ \overline{N}_{2R}
         \end{array}\right) 
 -{1\over 2}{\cal M}_{2I}^2
 (\overline{N}_{2I})^2,
\end{eqnarray}
with the mass matrices 
\begin{eqnarray}
{\cal M}_{2I}^2
&=& {1 \over 2f_{22}} 
{\partial^2 V\over \partial N_2^\dagger \partial N_2} 
\\
  {\cal M}^2&\equiv&
  \left(\begin{array}{cc}
   {1\over \sqrt{\lambda_+}}& \\
    &{1\over \sqrt{\lambda_-}}\\
        \end{array}\right)
   \left(\begin{array}{cc}
    \cos\theta&\sin\theta \\
    -\sin\theta &\cos\theta\\
         \end{array}\right)
  \left(\begin{array}{cc}
    {\partial^2 V\over \partial R^2 } &
    {\partial^2 V\over \partial R \partial N_2} \\
    {\partial^2 V\over \partial N_2^\dagger \partial R}
    & {\partial^2 V\over \partial N_2^\dagger \partial N_2} \\
         \end{array}\right)
\nonumber
\\
 &&\qquad\qquad
 \times
 \left(\begin{array}{cc}
    \cos\theta&-\sin\theta \\
    \sin\theta &\cos\theta\\
         \end{array}\right)
\left(\begin{array}{cc}
   {1\over \sqrt{\lambda_+}}& \\
    &{1\over \sqrt{\lambda_-}}\\
        \end{array}\right) .
\end{eqnarray}
where the potential $V$ is given in Eq.(\ref{potentialwp0}). 
For $c=c_{\textrm{\scriptsize cr}}-\Delta c$ with a small 
$\Delta c$, the mass squared  matrix 
at the leading order of $e^{-Rk\pi}$
is obtained 
as 
\begin{eqnarray}
 {\cal M}^2
  \approx k^2 w_0^2
  \left(\begin{array}{cc}
     {2(1-2c)^2\over 3-2c}(Rk\pi)^2 e^{(-4c^2+12c-10)Rk\pi}&
 {(2c-1)^{5/2}\over (3-2c)^2}(Rk\pi) e^{(-4c^2+11c-17/2)Rk\pi}  
 \\
   {(2c-1)^{5/2}\over (3-2c)^2}(Rk\pi) e^{(-4c^2+11c-17/2)Rk\pi}
  &{2c-1\over 4}(-4c^2+12c-6+{4(3-2c)\over 3})
e^{(2c-3)Rk\pi} 
        \end{array}\right) 
\nonumber
\end{eqnarray}
In this approximation, the mass squared of the imaginary 
part of the moduli field $\overline{N}_{2I}$ is the same 
as the right-bottom component of this mass squared matrix. 
Diagonalizing the mass squared matrix and 
using Eqs.(\ref{stanrk}) and (\ref{Rdeltac}),
we find that  the lighter physical mode is almost exclusively 
made of the radion 
\begin{eqnarray}
 m^2_{\textrm{\scriptsize light
}}
 &\approx& k^2 w_0^2 
   \frac{2(1-2c)^2}{3-2c}(Rk\pi)^2 e^{(-4c^2+12c-10)Rk\pi} 
\nonumber
\\
   &\approx& k^2 w_0^2 
  0.38 (3.4+ \ln \Delta c)^2 (\Delta c)^{1.7} . 
\end{eqnarray}
The heavier eigenmode is found to be exclusively made of 
the real part of moduli field 
\begin{eqnarray}
 m^2_{\textrm{\scriptsize heavy
}}
 &\approx& k^2 w_0^2 
 {(2c-1)\over 4}[-4c^2+12c-6+\frac{4}{3}(3-2c)]
{e^{(2c-3)Rk\pi} \over 1-e^{-(2c-1)Rk\pi}} 
\nonumber
\\
   &\approx& k^2 w_0^2 
  0.47 (\Delta c)^{0.70} . 
\end{eqnarray}
The imaginary part of the moduli field has the same mass as 
the real part of the moduli field in this approximation. 

For $\Delta c < 10^{-6}$ corresponding to $Rk> 1$, 
the physical mass of the light field (radion) is given by 
\begin{eqnarray}
   m_{\textrm{\scriptsize light}} < k w_0\times 
    10^{-4} . 
\end{eqnarray}
The physical mass of the heavy field (complex moduli) is given by 
\begin{eqnarray}
   m_{\textrm{\scriptsize heavy}} < 
 k w_0 \times 10^{-2}.  
\end{eqnarray}
We estimate the mass of the lighter physical mode (almost 
exclusively made of the radion), and that of the heavier 
mode (almost exclusively made of the 
complex moduli field) as 
\begin{eqnarray}
  m_{\textrm{\scriptsize light}} \sim  1 \textrm{TeV}, 
\qquad 
  m_{\textrm{\scriptsize heavy}} \sim 
 100 \textrm{TeV}
\end{eqnarray}
for $w_0\sim (10^{7}\textrm{GeV}/k)$ and $\Delta c\sim 10^{-6}$.

\setcounter{equation}{0}
\section{Soft mass by anomaly mediation}\label{sc:08082006-3anomaly}

In this section, we calculate soft masses
induced by anomaly mediation
in our hyperscalar background. 
In a supersymmetric Randall-Sundrum model,
anomaly-mediated scalar mass is given by
\cite{Luty:2002ff}
\begin{eqnarray}
  \tilde{m}_{\textrm{\scriptsize AMSB}}
 \sim {g^2\over 16\pi^2} \left\langle{F_\omega\over \omega}\right\rangle
\end{eqnarray}
where the superfield $\omega$ is defined as 
a rescaled compensator multiplet $\omega=\varphi e^{-T\sigma}$
and we denoted its lowest component also as $\omega$, 
and $g$ is gauge coupling constant for visible sector fields. 
In our model, the anomaly-mediated scalar mass becomes
\begin{eqnarray}
  \tilde{m}_{\textrm{\scriptsize AMSB}}
 \sim {g^2\over 16\pi^2}  (F_\varphi-F_T\sigma)\bigg|_{y=\pi}
\end{eqnarray}
The relevant F component is 
\begin{eqnarray}
F_{\varphi}-F_T\sigma &\!\!\!=&\!\!\! -\frac{e^{-R\sigma}}{R}
\left[
-\frac{1}{6M_5^3}\phi^\dag \partial_y \phi^{c\dag} 
-\frac{1}{3M_5^3}\phi^{c\dag}\partial_y \phi^\dag 
+\frac{1}{6M_5^3} \phi^\dag \phi^{c\dag}
\left( \frac{9}{2} -c \right)R\sigma'
-\frac{1}{2M_5^3}W_b 
\right. \nonumber \\
&\!\!\!&\!\!\! \left.
-\frac{3}{r}\phi^{c\dag} 
\partial_y \phi^\dag -\frac{3}{r}W_b 
+\frac{1}{r}
\phi^{c\dag}\phi^\dag \left(\frac{3}{2} -c \right)R\sigma'
\right] \label{f-f}
\end{eqnarray}
With background solutions (\ref{phin2}) and (\ref{phichic}),
Eq.(\ref{f-f}) at the boundary $y=\pi$ becomes
\begin{eqnarray}
  (F_\varphi-F_T\sigma)\bigg|_{y=\pi}
 &=&\hat{\epsilon}\sigma'e^{-Rk\pi}{\hat{N}e^{3Rk\pi} w_0 \over 
     \hat{N}(e^{2Rk\pi}-1)+{2\over 1-2c}(e^{-(1-2c)Rk\pi}-1)}
\nonumber
\\
 &\approx& - \hat{\epsilon}\sigma' w_0
  {2c_\textrm{\scriptsize cr}-1\over
   3-2c_\textrm{\scriptsize cr}}
  \sim 
  - \hat{\epsilon}\sigma' w_0 \times 0.05 .
\end{eqnarray}
where we used the stationary condition (\ref{stanrk}) in the second
equality.
Therefore we obtain the anomaly-mediated scalar mass as
\begin{eqnarray}
 \tilde{m}_{\textrm{\scriptsize AMSB}}
  \sim  {\cal O}(10^{-4})\times g^2 k w_0
\end{eqnarray}
As shown in the previous section, 
a lighter physical mass among the radion and moduli
can be of the order of a TeV
for $w_0\sim (10^{7}\textrm{GeV}/k)$.
For $g^2 kw_0\sim 10^6$GeV,
we obtain
\begin{eqnarray}
   \tilde{m}_{\textrm{\scriptsize AMSB}}
 \sim  100 \textrm{GeV} ,
\end{eqnarray}
which is a typical soft mass.

For gaugino mass, anomaly mediation is also 
dominant as long as additional interactions with gauge singlets
are not included.
The gaugino mass is of the same order as the scalar mass.

\setcounter{equation}{0}
\section{Soft mass by bulk field mediation}
\label{sc:w0model-4}

In this section, 
we examine soft mass induced by mediation of bulk fields.
In order to perform this computation, we need Kaluza-Klein mass spectrum
of all bulk fields.
Since the constant superpotential here only affect  
hyperscalar and gravitino,
what we have to do is to calculate the mass spectrum 
of these two fields.
The mass spectrum of the other fields has been known already and 
can be read from that of hyperscalar and gravitino as the
$w_0\to 0$ limit.
By using these resulting mass spectrum, 
we calculate  soft mass induced by mediation of bulk fields.

\subsection{Kaluza-Klein masses of hyperscalar}

Let us calculate mass spectrum of hyperscalar.
The analysis is similar to Ref.\cite{Maru:2006id}.
However we should carefully adopt approximations 
to use asymptotic forms of higher transcendental functions
for $\Delta c\sim 10^{-6}$.
In case where $\Delta c\sim 10^{-6}$ corresponding to
almost no exponential suppression
$e^{-Rk\pi}\sim e^{-\pi}\sim 0.04$,
there are two candidates of approximation:
\begin{eqnarray}
 \textrm{(I)} &&
  m_n/ k \ll 1  \textrm{~and~}
   m_n e^{kR\pi}/ k \gg 1 ,
\label{eq:approx1}
\\
 \textrm{(II)} &&
    m_n/ k \gg 1  \textrm{~and~}
   m_n e^{kR\pi}/ k \gg 1 .
\label{eq:approx2}
\end{eqnarray}
The approximation (I) is used also in our previous 
work in Ref.\cite{Maru:2006id}, 
where we have studied only the case with $w_\pi\not=0, w_0=0$ 
in detail. 
We will use the approximation (I) in the present model 
of $w_0\not=0, w_\pi=0$ in order to obtain the excitation 
spectra for mass much smaller compared to the mass scale 
$k$ of the AdS space. 
On the other hand,
the approximation (II) is more appropriate 
in $\Delta c\sim 10^{-6}$ model
since it is usable for most of $n$. 
Then 
eigenfunctions are found to behave as cosine or sine 
similarity to the flat case, as easily derived 
from properties of Bessel functions. 

We will perform an analysis similar to Ref.\cite{Maru:2006id} 
with both approximations (I) and (II).
The detail of calculations is shown in Appendix~\ref{ap:w0model-4}.
For the approximation (I), we find the hyperscalar mass  
\begin{eqnarray}
 m_n=ke^{-Rk\pi}\times\left\{
  \begin{array}{l}
  \left(n+{c+1\over 2}\right)\pi
   +{c(1-c)\over 2n\pi}\left(
     1+ 10^{-3}\left(
     {n\pi\over w_0}e^{-Rk\pi}\right)^2\right) ,
 \\
 \left(n+{c+1\over 2}\right)\pi   
     - {2n\pi\over c(c-1)}10^{-3}\left(
     {n\pi\over w_0}e^{-Rk\pi}\right)^2 ,
 \\
  \end{array}\right. \label{hya1}
\end{eqnarray}
where $c=c_{\textrm{\scriptsize cr}}-\Delta c$ and
$\Delta c\sim 10^{-6}$.
As expected, the $w_0$-dependent terms  
are highly suppressed by exponential factors.
For the approximation (II), we find the hyperscalar mass 
\begin{eqnarray}
 m_n\approx {k\over e^{Rk\pi}-1}\times
  \left( n\pi \pm \frac{w_0}{2\sqrt{3}}\right) . \label{hya2}
\end{eqnarray}
In case where $w_0\ll 1$,
perturbative treatment in Eq.(\ref{hya1})
seems to be broken.
Then only Eq.(\ref{hya2}) gives a valid representation 
for the hyperscalar mass.

\subsection{Kaluza-Klein masses of gravitino}\label{sc:gravitino-2}

Let us calculate mass spectrum of gravitino which is
the other superparticle affected by $w_0$. 
The relevant gravitino Lagrangian in the bulk is given 
by\footnote{Our convention is compatible with 
Refs.\cite{Wess-Bagger,Hebecker:2001ke}
and is summarized in Appendix~\ref{ap:gravitino-2}.}
\cite{Gherghetta:2000qt}
\begin{eqnarray}
{\cal L}_{{\rm bulk}} &=& 
 M_5\sqrt{-g}  \left[i\bar{\Psi}_M^i 
\gamma^{MNP} D_N \Psi_P^i -\frac{3}{2} \sigma' 
\bar{\Psi}_M^i \gamma^{MN}(\sigma_3)^{ij} \Psi_N^j \right] ,
 \label{ginolag}
\end{eqnarray}
\begin{eqnarray}
\Psi_M^1 &=& (\psi_M^1{}_{\alpha}, 
            \bar{\psi}_M^2{}^{\dot{\alpha}})^T, ~~
\Psi_M^2 = (\psi_M^2{}_{\alpha}, 
            -\bar{\psi}_M^1{}^{\dot{\alpha}})^T, \\
D_M &=& \partial_M + \omega_M, ~~
\omega_M =(\omega_\mu, \omega_4) = (\sigma' \gamma_4 \gamma_\mu/2, 0), 
 \\
\gamma^{M_1 M_2\cdots M_N} &=&
  \gamma^{[ M_1} \gamma^{M_2} \cdots \gamma^{M_N]} 
\nonumber \\
 &\equiv& \frac{1}{N!}(\gamma^{M_1} \gamma^{M_2} \cdots \gamma^{M_N} 
   + \textrm{antisymmetric permutations}), 
\end{eqnarray}
where the 5D curved indices are labelled by $M, N = 0, 1,2, 3, 4$.
The gamma matrix with curved indices is defined
through 5D vielbein as $\gamma^M=e_A^M\gamma^A$, 
where $A$ denote tangent space indices. 
In the second term in Eq.(\ref{ginolag}), 
$SU(2)_R$ indices are contracted by $(\sigma_3)$.

Boundary terms for gravitino are also contained 
in the term with the boundary superpotential $W_b$ 
in the superfield Lagrangian in Eq.(\ref{lagrangian}). 
By restoring the fermionic part, we find \cite{Kugo:2002js}
\begin{eqnarray}
{\cal L}_{\textrm{\scriptsize bound sup}} 
= \int d^2\theta \varphi^3 e^{-3T\sigma} W_b
= 3 \left[F_\varphi - \frac{1}{M_5^2}
\psi^1_\mu\sigma^{[\mu}\bar \sigma^{\nu]}\psi^1_\nu 
+ h.c.\right] W_b + \cdots
 ,
  \label{eq:sup_bound_pot} 
\end{eqnarray}
using $\varphi=1+\theta^2F_\varphi$. 
Therefore we obtain a boundary mass term for gravitino 
associated to the boundary superpotential 
\begin{eqnarray}
{\cal L}_{{\rm boundary}} = 
 -\frac{3W_b}{M_5^2} 
  \left[\psi_{\mu}^1 \sigma^{[\mu}\bar{\sigma}^{\nu]} \psi_\nu^1 
      +\bar{\psi}_{\mu}^1 \bar{\sigma}^{[\mu}\sigma^{\nu]}
     \bar{\psi}_\nu^1 \right] ,
  \label{ginolagb} 
\end{eqnarray}
where $W_b$ is  the constant superpotential localized at $y=0$   
given in Eq.(\ref{eq:boundary_pot}) and 
we assumed the $Z_2$ parity of $\psi_\mu^{1(2)}$ to be even (odd). 
From the Lagrangian given above,
we calculate mass spectrum.
The detail is shown in Appendix~\ref{ap:gravitino-2}.

For the lightest mode, we consider the limit 
\begin{equation}
\frac{m_n}{k} \ll 1, ~~\frac{m_n}{k}e^{Rk\pi} \ll 1. 
\end{equation}
In this limit, we find 
\begin{eqnarray}
  m_{\textrm{\scriptsize lightest}} \approx 6w_0 k ,
  \label{gino}
\end{eqnarray}
which can be $10^7$GeV for $w_0 \sim (10^7\textrm{GeV}/k)$.
This shows that the 4D gravitino (lightest mode) 
is much heavier than the 
the radion as well as scalars of the visible sector.
This is similar to the supersymmetry-breaking mediation model 
considered previously by Ref.\cite{Luty:2002ff}. 

For heavier KK modes of gravitino, 
we consider two limits
\begin{eqnarray}
&&\textrm{(I)}~~~\frac{m_n}{k} \ll 1, \frac{m_n}{k}e^{Rk\pi}\gg 1, \\
&&\textrm{(II)}~~~\frac{m_n}{k} \gg 1, \frac{m_n}{k}e^{Rk\pi}\gg 1. 
\end{eqnarray}
In the limit (I), we find 
\begin{eqnarray}
  m_n \approx 6w_0 k, ~
\left( n + \frac{1}{4} \right)\pi ke^{-Rk\pi} 
  \label{ginoh1}
\end{eqnarray}
where $n$ is an integer satisfying 
$\left( n + \frac{1}{4} \right)\pi e^{-Rk\pi} \ll 1$. 
The former one is the lightest mode solved above. 
In the limit (II), 
we find the mass
\begin{eqnarray}
 m_n \approx \left( n - \frac{6w_0}{2\pi} \right)\pi k e^{-Rk\pi}
  \label{ginoh2}
\end{eqnarray}
where $n$ is an integer satisfying 
$\left( n - \frac{6w_0}{2\pi} \right)\pi e^{-Rk\pi} \gg 1$.

\subsection{Induced masses by Kaluza-Klein modes}

In general, scalars in the visible sector 
can receive masses due to supersymmetry-breaking effects by 
mediation of bulk fields or   
by mediation of all modes in Kaluza-Klein decomposition. 
For example, take masses of bosons and fermions to be 
$n/R$ or $(n+1/2)/R$, where 
the mass splitting is $1/(2R)$.
This type of mass splitting has been considered also in 
the context of string theory 
\cite{Antoniadis:1997ic,Randall:1998uk}.
In these models,
it is known that
induced soft mass is of the order of%
\begin{eqnarray}
 \tilde{m}_{\textrm{\scriptsize KK-med}}
  \sim 10^{-1}\times  {m_{3/2}^2\over M_4} . 
 \label{anto}
\end{eqnarray}

The constant superpotential $w_0$ at $y=0$ 
induces soft mass
by mediation of all Kaluza-Klein modes.
Although we might be able to explicitly calculate 
the induced mass as in \cite{Antoniadis:1997ic,Randall:1998uk},
here we adopt a different way.
In our model the mass splitting between the bosons and fermions 
can be much smaller than $1/(2R)$ in the case above,
since the mass splittings are proportional to
positive powers of $w_0$ as seen from
Eqs.(\ref{hya2}) and (\ref{gino})-(\ref{ginoh2}).
Thus induced mass in our model should be small
compared to Eq.(\ref{anto}).
Thereby we can show that 
soft masses by mediation of Kaluza-Klein modes
in our model are smaller than
that of anomaly mediation.
The soft mass by mediation of Kaluza-Klein modes
is evaluated as
\begin{eqnarray}
  \tilde{m}_{\textrm{\scriptsize KK-med}}
   \lesssim 
 10^{-1}\times  {m_{\textrm{\scriptsize lightest}}^2\over M_4} 
  \sim 10^{-5} \textrm{GeV}
  \ll \tilde{m}_{\textrm{\scriptsize AMSB}} ,
\end{eqnarray}
where we used $m_{\textrm{\scriptsize lightest}}\sim 10^7\textrm{GeV}$
given below Eq.(\ref{gino}).
Therefore our model passes the FCNC constraint
also with respect to 
bulk field mediation while 
$\tilde{m}_{\textrm{\scriptsize AMSB}}\sim 100\textrm{GeV}$.

\setcounter{equation}{0}
\section{Cancellation of the cosmological constant
 by $F_X$
\label{sc:08082006-3}}
As mentioned in Sec.\ref{sc:model},
the potential in our model is negative
at the stationary point.
Nevertheless we have analyzed assuming 4D flat background. 
In order for our analysis to be consistent,
we should check whether additional sources can cancel the 
cosmological constant. 
As candidates of such sources, we consider an F term 
contribution and a D term contribution. 
In this section, we examine cancellation of the 
cosmological constant by a localized F term and check 
whether the induced soft mass is small compared to that 
of anomaly mediation. 
In the next section, we will examine the cancellation 
of the cosmological constant by the Fayet-Iliopoulos D 
term. 

We consider the setup where the hidden sector spurion%
\footnote{
Here we assume that the vacuum expectation value of the 
scalar component of $X$ vanishes: $\langle X \rangle = 0$. 
If $X$ is dynamical, one might expect 
that Coleman-Weinberg potential could generate nonzero $\langle X\rangle$.
%
A possible radiatively-induced K\"{a}hler potential 
$-(X^\dag X/(4\pi M_X))^2$ 
could indeed give a negative contribution to the
mass-squared of $X$ as $-(|F_X|/(4\pi M_X))^2 X^\dag X$,
where $M_X$ is the mass of $X$.
As we will see in (\ref{1010}), in the present model 
$F_X\sim (10^{10}\textrm{GeV})^2$. 
Our assumption $\langle X\rangle=0$ can be supported by
a positive mass-squared term $M_X^2 X^\dag X$ for $X$ with 
$M_X \gtrsim 10^{10}$GeV.

} chiral multiplet $X$ 
is localized at $y=0$ and the visible sector is localized at $y=\pi$. 
The Lagrangian is given by
\begin{eqnarray}
  {\cal L}_X = \left[
  \int d^4\theta \varphi^\dagger \varphi X^\dagger X
   + \int d^2\theta (\varphi^3 m^2 X + \textrm{h.c.})\right] \delta(y) ,
   \label{lagx}
\end{eqnarray}
with 
\begin{eqnarray}
    X=F_X \theta^2 ,
\end{eqnarray}
where $m$ is a mass parameter.
The auxiliary field Lagrangian of (\ref{lagx}) is
\begin{eqnarray}
  {\cal L}_X{}_{\textrm{\scriptsize aug}}
  = \left[ |F_X|^2 +(m^2 F_X + \textrm{h.c.})
  \right]  \delta(y) .
 \label{lagxa}
\end{eqnarray}
After solving equations of motion for $F_X$ from this equation
and performing $y$-integral,
we obtain
\begin{eqnarray}
   {\cal L}_X{}_{\textrm{\scriptsize aug}}
  = -m^4 = -|F_X|^2.
\end{eqnarray}
As seen from (\ref{lagxa}), the F-component of the spurion 
field does not mix with the F-components of 
$\Phi, \Phi^c,\varphi, T$. 
Therefore it does not change spectrum of 
$\Phi, \Phi^c,\varphi, T$ at tree level. 
However, at loop level it can contribute to visible sector soft masses.
In this case, 
the gravity multiplet can transmit supersymmetry breaking
to the visible sector at one-loop (which is 
called as brane-to-brane mediation 
by gravity). 
According to Ref.\cite{Gregoire:2004nn}, 
the induced soft mass is given by 
\begin{eqnarray}
\Delta \tilde{m}^2_{\textrm{\scriptsize bbg}} 
&=& -\frac{c_wk^4}{18 \pi^2 M_5^6} e^{-4k \pi R} |F_X|^2 
= -\frac{c_wk^2}{18 \pi^2 M_4^4} e^{-4k \pi R} |F_X|^2 , 
\end{eqnarray}
where 
$M_5^3 \simeq k M_4^2$ is used in the second equality 
and the dimensionless coefficient is of order unity: 
$c_w\sim {\cal O}(1)$. 
This mass should be suppressed compared 
to the anomaly-mediated scalar mass because it is tachyonic, 
\begin{eqnarray}
\frac{\Delta \tilde{m}^2_{\textrm{\scriptsize bbg}}}
 {\tilde{m}^2_{\textrm{\scriptsize AMSB}}} 
\sim -\frac{c_w}{18\pi^2}\omega_0^4 
\frac{|F_X|^2}{\tilde{m}^2_{\textrm{\scriptsize AMSB}}M_4^2}, 
\quad 
\left|\frac{\Delta \tilde{m}^2_{\textrm{\scriptsize bbg}}}
 {\tilde{m}^2_{\textrm{\scriptsize AMSB}}} \right| < 10^{-2} .
\end{eqnarray}
This gives a constraint to $F_X$,
\begin{eqnarray}
 F_X < 3\sqrt{2}\pi \tilde{m}_{\textrm{\scriptsize AMSB}} M_4
  \omega_0^{-2} \times 10^{-1}
\sim 10^{22}\textrm{GeV}^2  
  ~~~ \textrm{for}~Rk \sim 1 ,
\label{bbg}
\end{eqnarray}
where $\tilde{m}_{\textrm{\scriptsize AMSB}}\sim 100$GeV and
$\omega_0=e^{-Rk\pi}$ are used.

On the other hand, the contribution of the brane-to-brane mediation 
by hypermultiplet can 
be estimated following Ref.\cite{Maru:2003mq},
\begin{eqnarray}
\Delta \tilde{m}^2_{{\rm bbh}} 
&\sim& \frac{c_{ij}}{16\pi^2}\left( \frac{F_X}{\sqrt{3} M_4} \right)^2 
\left( \frac{k}{M_4} \right)^2
\left(\frac{1-2c}{e^{(1-2c)Rk\pi} -1 } \right)^2 e^{(3-2c)Rk\pi}
\end{eqnarray}
where $c_{ij}\sim {\cal O}(1)$.
This contribution should be also suppressed because this is 
flavor-violating, 
\begin{eqnarray}
\frac{\Delta \tilde{m}^2_{{\rm bbh}}}{\tilde{m}^2_{{\rm AMSB}}} 
&\sim& 
\frac{c_{ij}}{16\pi^2}\left(\frac{F_X}{\sqrt{3}M_4} \right)^2 
\left(\frac{k}{M_4} \right)^2
\left(\frac{2c-1}{e^{(1-2c)Rk\pi}-1} \right)^2 {e^{(3-2c)Rk\pi}
 \over\tilde{m}^2_{\rm AMSB}} < 10^{-3}  .
\end{eqnarray}
This leads to the condition
\begin{eqnarray}
F_X < {4\sqrt{3}\pi \over 10^3}
 M_4 \left( \frac{M_4}{k} \right) 
{1-e^{(1-2c)Rk\pi} \over (2c-1)}e^{-{1\over 2}(3-2c)Rk\pi} 
  \tilde{m}_{{\rm AMSB}}  
\sim 10^{18} \left(\frac{10^{18}}{k} \right)\textrm{GeV}^2 ,
\label{bbh}
\end{eqnarray}
for $Rk\sim 1$.
Taking into account the upper bound for AdS curvature scale 
$k < M_4 = 10^{18}$GeV, 
we find that the constraints (\ref{bbg}) and (\ref{bbh}) 
are satisfied by 
$\sqrt{F_X} \lesssim 10^{11}$ GeV. 

From (\ref{eq:pst}) we 
estimate 
the size of $\sqrt{F_X}$ to cancel the negative 
cosmological constant.
The size for $kw_0\sim 10^7$GeV
is obtained as
\begin{eqnarray}
  \sqrt{F_X} &\approx&  10^{12}  
  (\Delta c)^{0.3} ~ \textrm{GeV}
\nonumber
\\
  &\approx& 10^{10} 
  ~\textrm{GeV} 
  ~~~ \textrm{for}~~ \Delta c\sim 10^{-6} .
 \label{1010}
\end{eqnarray}
This shows that the cancellation of the cosmological constant 
can occur below (although near) the critical point for the 
FCNC constraint.

The contribution of $F_X$ to gravitino mass 
(irrespective of the value of $\langle X\rangle$)
is 
\begin{eqnarray}
 m_{3/2} &\sim& {F_X\over M_4}
\nonumber
\\
   &\approx& 100
  ~\textrm{GeV} 
 \label{m32f}
\end{eqnarray}
which is estimated based on effective 4D supergravity
below the compactification scale.
This is small compared to the contribution 
from the constant superpotential (\ref{gino}).

\setcounter{equation}{0}
\section{Cancellation of the 
cosmological constant by Fayet-Iliopoulos sector\label{sc:FIs}}
In this section, we examine a possibility of cancellation
of the cosmological constant 
by considering coupling of
Fayet-Iliopoulos sector localized at $y=0$.

We begin with the following Lagrangian as a simple extension of 
four-dimensional Fayet-Iliopoulos  model%
\footnote{%
Fayet-Iliopoulos terms have been related closely to R symmetry 
in the context of supergravity
\cite{Binetruy:2004hh,Abe:2004yk,VanProeyen:2004xt,Correia:2004pz,Abe:2004nx,Abe:2006eg}.
As an aspect of supergravity
it is straightforward to see 
from $\kappa \ll M_5$ that our model does not suffer from
Einstein term induced on brane by D term.
The authors thank Hiroyuki Abe for a valuable comment on this point.}
with the compensator
\begin{eqnarray}
 {\cal L}_V &=& \delta(y) \left[
\int d^4\theta \varphi^\dag \varphi
(A_1^\dag  e^{eV} A_1 + A_2^\dag e^{-eV} A_2 + 2\kappa V)
 \right.
\nonumber
\\
&&\qquad
 +\left. \left\{
\int d^2 \theta  
  \left(\textrm{${1\over 4}$}(W^{\alpha}W_{\alpha}+
   \bar{W}_{\dot{\alpha}}\bar{W}^{\dot{\alpha}})+
  \varphi^3( W(A_1, A_2) +{\rm h.c.})\right)
\right\}
\right] ,
 \label{FI}
\end{eqnarray}
with the superpotential
\begin{eqnarray}
  W(A_1,A_2)=m A_1 A_2 ,
\end{eqnarray}
where $A_1, A_2$ are charged chiral superfields, $V$ is a 
$U(1)$ vector superfield, and $W^\alpha$ is its field strength 
chiral superfield. 
The gauge coupling, Fayet-Iliopoulos parameter, 
and the $A_1, A_2$ mass parameter are 
denoted as $e$, $\kappa$ and $m$ respectively. 
The part of the Lagrangian containing auxiliary fields 
of this model reads 
\begin{eqnarray}
 {\cal L}_{V_{{\rm aux}}} &=&
\delta(y)
\left[ \textrm{${1\over 2}$}D^2+\kappa D +{e\over 2}
  (A_1^* A_1-A_2^* A_2) D
 \right.
 \nonumber
\\
 &&+ \left.
|F_\varphi|^2 (|A_1|^2 + |A_2|^2)
+|F_1|^2 + |F_2|^2 \right. \nonumber \\
&& \left. + \left\{
F_\varphi (A_1 F_1^* + A_2 F_2^*)
+3F_\varphi W
+ \frac{\partial W}{\partial A_1} F_1
+ \frac{\partial W}{\partial A_2} F_2
+{\rm h.c.}
\right\}
\right] \nonumber \\
\label{auxfi}
\end{eqnarray}
The equations of motion for $D$, $F_{1,2}$ are solved as
\begin{eqnarray}
0 &=& \frac{\partial {\cal L}_V{}_{{\rm aux}}}{\partial D}
= D+\kappa+{e\over 2}(A_1^* A_1-A_2^*A_2) , \\
0 &=& \frac{\partial {\cal L}_V{}_{{\rm aux}}}{\partial F_1^*}
= F_\varphi A_1 + F_1 + \left(\frac{\partial W}{\partial A_1} \right)^*, \\
0 &=& \frac{\partial {\cal L}_V{}_{{\rm aux}}}{\partial F_2^*}
= F_\varphi A_2 + F_2 + \left(\frac{\partial W}{\partial A_2} \right)^*
\end{eqnarray}
Putting these equations into (\ref{auxfi}) gives
\begin{eqnarray}
\Delta{\cal L} &=& \delta(y)
\left[ -\textrm{${1\over 2}$}D^2
- \left|\frac{\partial W}{\partial A_1} \right|^2
- \left|\frac{\partial W}{\partial A_2} \right|^2
 \right. \nonumber \\
&& \left. + \left\{ F_\varphi
\left(
3W - A_1\frac{\partial W}{\partial A_1}
- A_2\frac{\partial W}{\partial A_2}
\right) +{{\rm h.c.}}
\right\}
\right]
\end{eqnarray}
which implies that if
\bea
0 &=& 3W - A_1\frac{\partial W}{\partial A_1}
- A_2\frac{\partial W}{\partial A_2}
 =m A_1 A_2
\label{decouplefi}
\eea
the hidden sector does not affect equations of motion for 
the compensator.

Next we have to examine whether the condition (\ref{decouplefi})
is compatible with the minimization conditions of the 
hidden sector potential.
The hidden sector potential is given by
\begin{eqnarray}
\Delta V 
= \delta(y)
 \left[ \textrm{${1\over 2}$}(\kappa+{e\over 2}
  (A_1^*A_1-A_2^* A_2))^2
  +|mA_1|^2 + |mA_2|^2
-\left\{
F_\varphi m A_1 A_2 + {\rm h.c.}\right\}
\right]
\end{eqnarray}
The minimization conditions of the potential are
\begin{eqnarray}
0 &=& \frac{\partial \Delta V}{\partial A_1^*}
= (\kappa+{e\over 2}(A_1^*A_1-A_2^*A_2))
  ({e\over  2}A_1) 
+|m|^2 A_1 - (F_\varphi m A_2)^*
\label{mini1fi}\\
0 &=& \frac{\partial \Delta V}{\partial A_2^*}
= (\kappa+{e\over 2}(A_1^*A_1-A_2^*A_2))
  (-{e\over  2}A_2) 
+|m|^2 A_2 - (F_\varphi m A_1)^*
\label{mini2fi}
\end{eqnarray}
We find that a solution is given by 
\begin{eqnarray}
  A_1 =A_2 =0 .
\end{eqnarray}
At this minimum, 
the value of the additional 
potential (after $y$-integration) is 
\begin{eqnarray}
  \Delta V = {1\over 2} \kappa^2 .
\end{eqnarray} 
This positive contribution can cancel the negative 
vacuum energy found in Eq.(\ref{eq:pst}), if 
$\sqrt{\kappa}\approx 10^{10}$GeV. 
If matter in visible sector is neutral under the $U(1)$,
additional contributions to scalar masses are not generated.
%
The contribution of this sector to gravitino mass
is small similarly to (\ref{m32f}).

\setcounter{equation}{0}
\section{Conclusion}\label{sc:conclusion}

We have studied
radion and moduli masses and
induced soft masses 
in the radius stabilization model 
proposed previously in Ref.\cite{Maru:2006id}.
We have simultaneously obtained supersymmetry breaking and 
radius stabilization, as well as
soft mass without FCNC problem,
large gravitino mass, 
and radion mass of the order of TeV.
These all give evidence that
our model is phenomenologically viable.

In the model  
the potential has negative value at the stationary point.
In order to cancel the negative cosmological constant,
we have introduced localized F term at $y=0$ 
as another source of supersymmetry-breaking
and analyzed in 4D flat space.
In our model there are four parameters:
5D Planck mass $M_5$, 5D curvature $k$,
bulk mass parameter for hypermultiplet $c$ and
constant boundary superpotential $w_0$
(up to visible sector gauge coupling constant $g$).
Among them, one of dimensionful quantities gives 
unit of mass dimension.
In numerically evaluating 
various masses,
we have chosen
$M_5\sim (M_4^2 k)^{1/3}$, 
$w_0\sim (10^7\textrm{GeV}/k)$ and
$c=c_\textrm{\scriptsize cr}-\Delta c$ 
where $c_\textrm{\scriptsize cr}\approx 0.546$ 
and $\Delta c\sim 10^{-6}$.
This $c$ corresponds to the stabilized radius $R\sim k^{-1}$.

The quantum fluctuation of 
radion mixes with complex moduli.
We have found that the lightest physical mode of the admixture has
the mass of the order of 1TeV.
Such a comparatively small radion mass appears as a common feature of
warped space model\cite{Kribs:2006mq} and
its value
is in experimentally allowed region\cite{Yao:2006px}.

We have also found that soft mass is of the order of 100GeV
and is generated by anomaly mediation.
Therefore there is no FCNC problem. 
The gravitino mass is found to be $10^7$GeV.
Such a large gravitino mass is similar to that of 
the supersymmetry-breaking
mediation scenario given in Ref.\cite{Luty:2002ff}.
We have found that the hyperscalar mass is of the order of $k$
and is much heavier than other fields.
Therefore the hyperscalar primarily acts as a part of 
the background configuration.

As for the cancellation of the cosmological constant,
the FCNC constraint for localized F term 
leads to 
$\sqrt{F_X} \lesssim 10^{11}$GeV.
The cosmological constant is cancelled
for 
$\sqrt{F_X} \approx 10^{10}$GeV.
This justifies
our analysis based on 4D flat background.  
Similarly in the scenario with D term contribution, 
the cosmological constant is cancelled
for Fayet-Iliopoulos parameter $\sqrt{\kappa}\approx 10^{10}$GeV.

There remains to be still examined some issues 
in our model.
We have not considered
the imaginary part for the lowest component
of the radion supermultiplet.
In the present model, the mass seems to vanish 
because the potential can be seen to include only real part of the 
radion supermultiplet.
However,
the mass would be induced by quantum loop effect 
similarly to the case of gauge-Higgs unification 
model~\cite{Hosotani:1983xw}.
If the induced mass is very small,
it would have to be examined whether the coupling 
with other particles can be very small as in the case of axion.


The issue of radius stabilization by Casimir energy 
\cite{Ponton:2001hq} also remains to be examined.
Analysis about this point will be given 
in a separate paper \cite{MSU}.

\vspace*{10mm}
\begin{center}
{\bf Acknowledgements}
\end{center}  
The authors thank Yutaka Sakamura for a useful discussion. 
This work is supported in part by Grant-in-Aid for Scientific 
Research from the Ministry of Education, Culture, Sports, 
Science and Technology, Japan No.17540237 (N.S.) and 
No.18204024 (N.M.~and N.S.). 
The work of N.U.~is supported by Bilateral exchange programme between 
the Academy of Finland and the Japanese Society for Promotion
of Science.

\begin{appendix}
\setcounter{equation}{0}
\section{Calculations of mass spectrum of hyperscalar}\label{ap:w0model-4}

In this appendix, we give the detail of calculations
of mass spectrum of hyperscalar.
Let us consider
$n$-th Kaluza-Klein effective field 
$\phi_n^I(x)$ with its mode functions $b_n^I(y)$ as $\phi(x, y)$ 
component and $b_n^{cI}(y)$ as $\phi^c(x, y)$ component 
\begin{eqnarray}
 \left(
  \begin{array}{c}
   \phi(x,y)\\ \phi^c(x,y)
  \end{array}\right)
  =\sum_{n}
\sum_{I=1,2}
 \phi_n^I(x) 
  \left(
   \begin{array}{c}
    b_n^I
(y)\\ \hat \epsilon(y)b_n^c{}^I
(y)
   \end{array}\right), 
\label{modeexpand}
\end{eqnarray}
where $I$ is the indices corresponding to the two independent 
mass eigenfunctions of effective fields. 

Assuming that the effective four-dimensional field 
$\phi_n(x)$ has mass $m_n$, 
we easily find solutions in the bulk in terms of the 
Bessel functions $J_\alpha, (J_\beta)$ and 
$Y_\alpha, (Y_\beta)$ 
\cite{Gherghetta:2000qt}
\bea
b_n(y) &=& \frac{e^{2R\sigma}}{N_n}
\left[
J_\alpha(m_ne^{R\sigma}/k) 
+ b_\alpha(m_n)Y_\alpha(m_ne^{R\sigma}/k)
\right] , 
\quad \alpha =|c+\frac{1}{2}| , 
   \label{bn} \\
b_n^c(y) &=& \frac{e^{2R\sigma}}{N_n^c}
\left[
J_{\beta}(m_n e^{R\sigma}/k) 
+ b_{\beta}(m_n)Y_{\beta}(m_n e^{R\sigma}/k)
\right], 
\quad \beta=|c-\frac{1}{2}| . 
   \label{eq:bnc} 
\eea
In the equations of motion, singular terms 
give the boundary conditions. 
The first boundary condition comes from 
$\delta^2$ terms 
\bea
\label{BC1}
0&=&-2 b_n^c(0) + w_0 b_n(0) , \\
0&=& b_n^c(\pi)  . 
\label{BC2}
\eea
The second boundary condition comes from $\delta$ function 
\bea
\label{BC3}
\!\!\!\!\!&&
0 = \frac{7}{3}w_0 b_n^c(0) 
-\frac{1}{3}w_0 
\left[
2b_n^c(0) +\frac{m_n}{k} \frac{1}{N_n^c}
\left\{ J_{\beta}'(m_n/k) 
+ b_{\beta}(m_n) Y_{\beta}'(m_n/k) \right\}
\right] \nonumber \\
\!\!\!\!\!&&-4b_n(0) 
+ 2 \left( \frac{3}{2}-c \right)b_n(0)
-\frac{2m_n}{k}
\left[ \frac{1}{N_n} 
\left\{ J_\alpha'(m_n/k) 
+ b_\alpha(m_n) Y_\alpha'(m_n/k) \right\} \right],
\\
\!\!\!\!\!&&0 = 
(1+2c)b_n(\pi)
   +{2m_n\over  k}
   \left[
   {e^{3Rk\pi}\over N_n}
  \{J_\alpha'(m_n e^{Rk\pi}/k)
   +b_\alpha(m_n) Y_\alpha'(m_n e^{Rk\pi}/k)\}\right].
\label{BC4}
\eea
with 
$J'(z)=dJ(z)/dz$. 

Let us first consider the approximation (I) in 
Eq.~(\ref{eq:approx1}) to solve
(\ref{BC1})-(\ref{BC4}).
From Eqs.(\ref{BC1}) and (\ref{BC3}),
we find
\begin{eqnarray}
b_\alpha(m_n) 
&\sim&-
 \frac{4 \beta}{(\beta+5)w_0 } 
\frac{N_n}{N_n^c}  
  \left( \frac{m_n}{2k} \right)^{\alpha+\beta}
  \frac{\pi}{\Gamma(\beta+1)\Gamma(\alpha)} ,
\\
 b_\beta(m_n) 
&\sim& -
  \frac{12\alpha}{(\beta+5)w_0 }
\left( \frac{m_n}{2k} \right)^{\alpha+\beta}
  \frac{\pi}{\Gamma(\alpha+1)\Gamma(\beta)}
{N_n^c\over N_n} .
\end{eqnarray}
where we used\footnote{When we consider supersymmetric 
limit $w_0\to 0$,
we should take $1+2c-2\alpha=0$ after $w_0\to 0$.} 
 $1+2c-2\alpha=0$ for 
$c=c_\textrm{\scriptsize cr}-\Delta c, \Delta c >0$ 
and for simplicity we ignored $w_0$-independent terms. 
To solve the remaining Eqs.(\ref{BC2}) and (\ref{BC4}), 
we use the asymptotic behavior of the Bessel functions for 
$|z|\gg 1$ 
\begin{eqnarray}
    J_\alpha(z)\sim \sqrt{2\over \pi z}
    \left(\cos\left(z-{2\alpha+1\over 4}\pi\right) 
 -{4\alpha^2-1\over 8z}\sin\left(z-{2\alpha+1\over 4}\pi\right)\right)
\\
   Y_\alpha(z)\sim \sqrt{2\over \pi z}
    \left(\sin\left(z-{2\alpha+1\over 4}\pi\right) 
+{4\alpha^2-1\over 8z}\cos\left(z-{2\alpha+1\over 4}\pi\right)\right) .
  \label{bessellarge}
\end{eqnarray}
After changing variables, 
\bea
\frac{m_n}{k}e^{Rk\pi} \equiv x, 
\qquad 
\frac{2\alpha+1}{4}\pi \equiv a, 
\qquad 
\frac{2\beta+1}{4}\pi \equiv b,
\eea
we can rewrite the boundary conditions at $y=\pi$ and 
obtain the mass eigenvalue equation similarly 
to Ref.\cite{Maru:2006id} 
\begin{eqnarray}
 \tan^2(x-a)-A(x)\tan(x-a) +B(x)=0
\end{eqnarray} 
with
\begin{eqnarray}
 &&A(x)={c(1-c)\over 2x}\left(1-B(x)\right)
\\
 &&B(x)=-{48\alpha\beta\over (\beta+5)^2w_0^2}
    \left({xe^{-Rk\pi}\over 2}\right)^{2(\alpha+\beta)}
   {\pi^2\over
 \Gamma(\alpha+1)\Gamma(\alpha)\Gamma(\beta+1)\Gamma(\beta)}  
\end{eqnarray}
where $\cos(x-a)\neq 0,x\neq 0$.
Solving this equation, we find the hyperscalar mass in 
Eq.(\ref{hya1}).

Next we consider the approximation (II) in Eq.(\ref{eq:approx2}) 
instead of Eq.(\ref{eq:approx1}) to solve 
(\ref{BC1})-(\ref{BC4}).
After substituting
Eq.(\ref{bessellarge}) into the boundary conditions,
we obtain the mass eigenvalue equation
\begin{eqnarray}
   \left(\tan[(1-e^{-Rk\pi})x]\right)^2 
  \approx \frac{w_0^2}{12}
  \label{sol}
\end{eqnarray}
By solving this equation, we find the hyperscalar mass 
in Eq.(\ref{hya2}).

\setcounter{equation}{0}
\section{Calculations of mass spectrum of gravitino}
\label{ap:gravitino-2}

In this appendix, we give our convention 
and some details of calculations
of mass spectrum of gravitino.

The gamma matrices are given by 
\cite{Wess-Bagger,Hebecker:2001ke,Mirabelli:1997aj}
\begin{eqnarray}
 \gamma^m =\left(
  \begin{array}{cc}
   0 &\sigma^m \\
   \bar{\sigma}^m &0 \\
  \end{array}\right)
\end{eqnarray}
where $m=0,1,2,3$, and
\begin{eqnarray}
 \gamma^4=\gamma^0\gamma^1\gamma^2\gamma^3
  =\left( \begin{array}{cc}
   -i & 0 \\
   0 &i \\
  \end{array}\right) .
\end{eqnarray}
We need to choose $\gamma^4$ to be anti-hermitian, since
\begin{eqnarray}
 \gamma^A \gamma^B +\gamma^B \gamma^A= -2\eta^{AB} ,\quad
  \eta^{AB}=(-1,+1,+1,+1,+1) .
\end{eqnarray}
We define hermitian chiral gamma matrix as
\begin{eqnarray}
 \gamma_5\equiv i\gamma^4
    =\left( \begin{array}{cc}
   1 & 0 \\
   0 &-1 \\
  \end{array}\right) .
\end{eqnarray}
Charge conjugation matrix is given by
\begin{eqnarray}
 C=\left( \begin{array}{cc}
   i\sigma^2 & 0 \\
   0 &i\sigma^2 \\
  \end{array}\right) 
 =\left( \begin{array}{cccc}
   0 &1&& \\
   -1&0&& \\
    &&0&1 \\
    &&-1&0\\
  \end{array}\right) 
 =\left( \begin{array}{cc}
   -\varepsilon_{\alpha\beta}& 0\\
   0&\varepsilon^{\dot{\alpha}\dot{\beta}}\\
  \end{array}\right) ,
\end{eqnarray}
where $\varepsilon_{12}=-1$, $\varepsilon^{12}=+1$ and
\begin{eqnarray}
 C \gamma^A C^{-1}=\gamma^A {}^T .
\end{eqnarray}
The four-component spinors $\Psi^1$ and $\Psi^2$ satisfy
symplectic Majorana condition
\begin{eqnarray}
 \Psi^i=\varepsilon^{ij}C\bar{\Psi}^j{}^T
\end{eqnarray}
where Dirac conjugate is defined as
\begin{eqnarray}
 \bar{\Psi}=\Psi^\dagger \gamma^0 .
\end{eqnarray}
The $Z_2$ projection is defined as 
\begin{eqnarray}
 \Psi_M^i(-y)=(\sigma^3)^{ij}\gamma_5 \Psi^j(y)
\end{eqnarray}
which can be rewritten in terms of two-component spinors as
\begin{eqnarray}
 \psi_M^1{}_{\alpha}(-y)
  &=&\psi_M^1{}_{\alpha}(y) ,\\
\psi_M^2{}_{\alpha}(-y)
  &=&-\psi_M^2{}_{\alpha}(y) .
\end{eqnarray}

We now give some details of calculations
of mass spectrum of gravitino. 
Since we are interested in 4D gravitino and its KK modes, 
we ignore the extra-dimensional component $\psi_y$. 
From the Lagrangian (\ref{ginolag}) and (\ref{ginolagb}),
the equations of motion for gravitino are
\begin{eqnarray}
0 &=& i\bar{\sigma}^{[\mu}\sigma^\nu\bar{\sigma}^{\rho]}
   \partial_\mu \psi_\nu^1 
  -\bar{\sigma}^{[\nu}\sigma^{\rho]}
  \left({1\over R}\partial_y-{3\over 2}\sigma'\right)
  \bar{\psi}_\nu^2
+\frac{6w_0}{R}\delta(y)
  \bar{\sigma}^{[\nu}\sigma^{\rho]}\bar{\psi}_\nu^1, 
\\
0 &=& i\sigma^{[\mu}\bar{\sigma}^\nu\sigma^{\rho]} 
    \partial_\mu \bar{\psi}_\nu^2 
  +\sigma^{[\nu}\bar{\sigma}^{\rho]}
    \left({1\over R}\partial_y+{3\over 2}\sigma'\right)
   \psi_\nu^1
\end{eqnarray}
which are further simplified 
to
\begin{eqnarray}
0 &=& -i\bar{\sigma}^{\mu}\partial_\mu \psi_\nu^1 
  +\left({1\over R}\partial_y-{3\over 2}\sigma'\right)
  \bar{\psi}_\nu^2
 -\frac{6w_0}{R}\delta(y)\bar{\psi}_\nu^1, 
\\
0 &=& -i\sigma^{\mu}\partial_\mu \bar{\psi}_\nu^2 
    -\left({1\over R}\partial_y+{3\over 2}\sigma'\right)
   \psi_\nu^1
\end{eqnarray}
up to the gauge fixing $\bar{\sigma}^\mu \psi_\mu^{1,2}=0$ and 
$\partial^\mu \psi^{1,2}_\mu =0$.

Let us take
mode expansions 
\begin{eqnarray}
\psi_\rho^{1,2}(x,y) = \sum_n \psi_\rho^{1,2(n)}(x) f^{1,2(n)}(y) ,
\end{eqnarray}
where the 4D effective fields $\psi_\rho^{1,2(n)}(x)$ have 
mass $m_n$ 
\begin{eqnarray}
-i\bar \sigma^m \partial_m \psi_\rho^{1(n)}=m_n \bar \psi_\rho^{2(n)}, 
\end{eqnarray}
\begin{eqnarray}
-i\sigma^m \partial_m \bar \psi_\rho^{2(n)}=m_n \bar \psi_\rho^{1(n)}. 
\end{eqnarray}
The solutions in the bulk are obtained as \cite{Gherghetta:2000qt}
\begin{eqnarray}
f_1^{(n)}(y) &=& \frac{e^{R\sigma/2}}{N_n} 
\left[ J_2\left(\frac{m_n}{k}e^{R\sigma} \right) 
+ b_2(m_n) Y_2 \left(\frac{m_n}{k} e^{R\sigma} \right) \right], \\
f_2^{(n)}(y) &=& \hat \epsilon(y) \frac{e^{R\sigma/2}}{N_n} 
\left[ J_1\left(\frac{m_n}{k}e^{R\sigma} \right) 
+ b_1(m_n) Y_1 \left(\frac{m_n}{k} e^{R\sigma} \right) \right]
\end{eqnarray}
where $\hat \epsilon(y)$ is defined in Eq.(\ref{eq:sign_function}).

The boundary conditions we should consider are 
\begin{eqnarray}
0 &=& \bar{\tilde{f}}_2^{(n)}(0) - {6w_0\over 2}\bar{f}_1^{(n)}(0),
  \label{bcg1}\\ 
0 &=& \tilde{f}_2^{(n)}(\pi),
  \label{bcg2} \\
0 &=& \left(\partial_y + \frac{3}{2} R\sigma' \right)f_1^{(n)}(\pi) ,
  \label{bcg3}
\end{eqnarray}
where $f_2^{(n)}(y) \equiv \hat \epsilon(y)\tilde{f}_2^{(n)}(y)$. 
Eq.(\ref{bcg1}) gives 
\begin{eqnarray}
0 &=&  \left[ J_1\left(\frac{m_n}{k} \right) 
+ b_1(m_n) Y_1 \left(\frac{m_n}{k} \right) \right] 
  - {6w_0\over  2} \left[ J_2\left(\frac{m_n}{k} \right) 
+ b_2(m_n) Y_2 \left(\frac{m_n}{k} \right) \right], 
\label{bc1gg} 
\end{eqnarray}
Eqs.(\ref{bcg2}) and (\ref{bcg3}) give 
\begin{eqnarray}
  b_1(m_n) = 
 -\frac{J_1(\frac{m_n}{k}e^{Rk\pi})}{Y_1(\frac{m_n}{k}e^{Rk\pi})}
  = b_2(m_n) .
  \label{bc2gg}
\end{eqnarray}

For the lightest mode, we consider the limit 
\begin{equation}
\frac{m_n}{k} \ll 1, \frac{m_n}{k}e^{Rk\pi} \ll 1. 
\end{equation}
Using the asymptotic form of Bessel functions for $|z|\ll 1$
and (\ref{bc2gg}), 
the boundary condition (\ref{bc1gg}) can be rewritten as 
\begin{eqnarray}
0 &\approx& \left[\frac{m_n}{2k} 
-\frac{\frac{m_n}{2k}e^{Rk\pi}}{-\frac{1}{\pi}
 \frac{2k}{m_n}e^{-Rk\pi}} 
\left(-\frac{1}{\pi} \right) \frac{2k}{m_n} \right]
\nonumber \\
&& -{6w_0\over 2}
\left[\frac{1}{2} \left(\frac{m_n}{2k} \right)^2 
-\frac{\frac{m_n}{2k}e^{Rk\pi}}{\left(-\frac{1}{\pi} \right)
\frac{2k}{m_n}e^{-Rk\pi}} \left(-\frac{1}{\pi} \right)
\left( \frac{2k}{m_n} \right)^2 \right], 
\end{eqnarray}
From this equation, we find the mass spectrum in Eq.(\ref{gino}). 

For heavier modes, 
we consider two limits
\begin{eqnarray}
&&\textrm{(I)}~~~\frac{m_n}{k} \ll 1, \frac{m_n}{k}e^{Rk\pi}\gg 1, 
\label{eq:limit1} \\
&&\textrm{(II)}~~~\frac{m_n}{k} \gg 1, \frac{m_n}{k}e^{Rk\pi}\gg 1. 
\label{eq:limit2}
\end{eqnarray}
Using the approximations of Bessel function for $|z| \gg 1$
and Eq.(\ref{bc2gg}), 
we can reduce Eq.(\ref{bc1gg}) in the limit (I) in 
Eq.(\ref{eq:limit1}) to 
\begin{eqnarray}
0 &\approx& \left[ \frac{m_n}{2k} 
-\cot \left(\frac{m_n}{k}e^{Rk\pi} -\frac{3}{4}\pi \right)
\left(-\frac{1}{\pi} \right) \frac{2k}{m_n} \right] \nonumber \\
&& -{6w_0\over 2}
\left[
\frac{1}{2}\left(\frac{m_n}{2k} \right)^2 
-\cot \left(\frac{m_n}{k}e^{Rk\pi} -\frac{3}{4}\pi \right)
\left(-\frac{1}{\pi} \right) \left( \frac{2k}{m_n} \right)^2 \right] .
\end{eqnarray}
From this equation, we find the mass spectrum in Eq.(\ref{ginoh1}). 

On the other hand, 
in the limit (II) in Eq.(\ref{eq:limit2}) 
the boundary condition (\ref{bc1gg}) 
reduces to 
\begin{eqnarray}
0 &\approx& \left[ 
\cos \left(\frac{m_n}{k} - \frac{3}{4}\pi \right) 
-\cot \left(\frac{m_n}{k}e^{Rk\pi} - \frac{3}{4}\pi \right) 
\sin \left(\frac{m_n}{k} - \frac{3}{4}\pi \right) 
\right] \nonumber \\
&& -{6w_0\over 2}
\left[
\cos \left(\frac{m_n}{k} - \frac{5}{4}\pi \right) 
-\cot \left(\frac{m_n}{k}e^{Rk\pi} - \frac{3}{4}\pi \right) 
\sin \left(\frac{m_n}{k} - \frac{5}{4}\pi \right) 
\right] .
\end{eqnarray}
Thus we find the mass spectrum in Eq.(\ref{ginoh2}). 
\end{appendix}

\vspace*{10mm}


\end{document}